\documentclass[twocolumn]{IEEEtran}

\usepackage{amsmath,epsfig,amssymb,verbatim,amsopn,cite,multirow}
\usepackage[usenames,dvipsnames]{color}
\usepackage{algorithm,algorithmic}
\usepackage{amsfonts}
\usepackage{amssymb}
\usepackage{epsfig}
\usepackage{epstopdf}
\usepackage{graphicx}
\usepackage{footnote}

\usepackage[margin=15mm,top=18mm,bottom=27mm]{geometry}

\usepackage[nodisplayskipstretch]{setspace}
\newtheorem{proposition}{Proposition}

\usepackage{amsmath,epsfig,amssymb,verbatim,amsopn,cite,subfigure,multirow}
\usepackage{balance}
\usepackage{cite}

\newcommand{\Ex}{\mathbb{E}}

\newcommand{\qa}{{\bf a}}

\newcommand{\qg}{{\bf g}}
\newcommand{\qh}{{\bf h}}

\newcommand{\qt}{{\bf t}}

\newcommand{\qv}{{\bf v}}

\newcommand{\qx}{{\bf x}}

\newcommand{\qI}{{\bf I}}

\newcommand{\sen}{\mathrm{Sen}}
\newcommand{\Com}{\mathrm{Com}}
\newcommand{\ave}{\mathrm{ave}}
\newcommand{\IR}{\mathrm{IR}}
\newcommand{\tmsen}{\qt_{m}^{\mathrm{Sen}}}
\newcommand{\tmkcom}{\qt_{mk}^{\mathrm{Com}}}
\newcommand{\tmkpcom}{\qt_{mk'}^{\mathrm{Com}}}
\newcommand{\M}{\mathcal{M}}
\newcommand{\K}{\mathcal{K}_d}

\newcommand{\SINR}{\mathrm{SINR}}
\newcommand{\SE}{\mathrm{SE}}

\newcommand{\aaa}{\qa}
\newcommand{\ETAmk}{\boldsymbol{\eta}^{\Com}}
\newcommand{\ETAm}{\boldsymbol{\eta}^{\sen}}

\newcommand{\Sn}{\sigma^2_n}
\newcommand{\etamk}{\eta_{mk}}
\newcommand{\etam}{\eta_{m}}
\newcommand{\gammk}{\gamma_{mk}}

\usepackage{algorithm,algorithmic}

\newcommand{\SEk}{\mathrm{SE}_{k}}

\usepackage{MnSymbol}
\begin{document}

\author{
Mohamed Elfiatoure,  Mohammadali Mohammadi, Hien Quoc Ngo, and Michail Matthaiou\\
\small{
Centre for Wireless Innovation (CWI), Queen's University Belfast, U.K.\\
Email:\{melfiatoure01, m.mohammadi, hien.ngo, m.matthaiou\}@qub.ac.uk}}\normalsize

\title{Cell-Free Massive MIMO for ISAC: Access Point Operation Mode Selection and Power Control}

\maketitle
\begin{abstract}
This paper considers a cell-free massive multiple-input multiple-output (MIMO) integrated sensing and communication (ISAC) system, where distributed MIMO access points (APs) are used to jointly serve the communication users and detect the presence of a single target. We investigate the problem of AP operation mode selection, wherein some APs are dedicated for downlink communication, while the remaining APs are used for sensing purposes.  Closed-form expressions for the individual spectral efficiency (SE) and mainlobe-to-average-sidelobe ratio (MASR) are derived, which are respectively utilized to assess the communication and sensing performances. Accordingly, a max–min fairness problem is formulated and solved, where the minimum SE of the users is maximized, subject to the per-AP power constraints as well as sensing MASR constraint. Our numerical results show that the proposed AP operation mode selection with power control can significantly improve the communication performance for given sensing requirements.

\let\thefootnote\relax\footnotetext{This project has received funding from the European Research Council (ERC) under the European Union's Horizon 2020 research and innovation programme (grant agreement No. 101001331). The work of H. Q. Ngo was supported by the U.K. Research and Innovation Future Leaders Fellowships under Grant MR/X010635/1.}
\end{abstract}

\vspace{-1.5em}
\section{Introduction}
ISAC has recently been  envisioned as a key enabling technology for future wireless networks, aiming to efficiently utilize the congested resources for both communication and sensing~\cite{Liu:JSAC:2022,Liu:Tut:2022}. The radar bands set aside for sensing can be harnessed for wireless communication operation, enabling the implementation of high data-rate applications. To unify the radar and communication operations, two well-known designs, namely separated and co-located systems, were introduced in~\cite{Liu:WCL:2017,Li:TAES:2017,Elfiatoure:JCIN:2023} and~\cite{Chiriyath:TCCN:2017,Liu:TWC:2018}, respectively. The former utilizes different devices, operating over the same frequency band, for communication and sensing, while in the latter a single device acts as radar and communication base station (BS) by simultaneously communicating with multiple downlink users and detecting radar targets.

The main driving force behind the transition from the separated design to a co-located design was to reduce the complexity induced by side-information exchange among the radar and communication devices~\cite{Liu:TWC:2018}. However, co-located design with a MIMO BS often suffers from a fairness problem, since the cell-boundary users are subject to inter-cell interference and significant power decay over long distances. The key feature of massive MIMO technology, i.e., inter/intra-cell interference suppression, revitalizes the interest towards separated design with multiple communication and radar devices to implement distributed ISAC architectures. In this context, cell-free massive MIMO with distributed MIMO APs can be exploited to support ISAC. In cell-free massive MIMO, all  users are coherently served by all APs over the same time-frequency band. Each AP is connected to a central processing unit (CPU) via fronthaul links, and the CPU is responsible for coordination~\cite{Matthaiou:COMMag:2021,Hien:cellfree}.

The integration of ISAC into cell-free massive MIMO networks, has been recently investigated in~\cite{Behdad:GC:2022,demirhan2023cell}. Specifically, Behdad \textit{et al.} \cite{Behdad:GC:2022} studied a cell-free massive MIMO ISAC system, consisting of a fixed number of transmit and receive APs. Users are served by the transmit APs and, at the same time, the transmitted signals are used for sensing to detect the presence of a target in a certain location. The reflected signals are received at the receive APs and then processed at the CPU. The authors proposed a power allocation algorithm to maximize the sensing signal-to-noise ratio (SNR) under signal-to-interference-plus-noise ratio (SINR) constraints at the user. Demirhan \textit{et al.} \cite{demirhan2023cell} studied the sensing and communication beamforming design problem in cell-free massive MIMO ISAC systems, where a joint beamforming design was proposed to maximize the sensing SNR, while satisfying the communication SINR constraints.

Different from the above-mentioned works~\cite{Behdad:GC:2022,demirhan2023cell}, where the AP operation modes are fixed, we consider a novel cell-free massive MIMO ISAC network with dynamic AP operation mode selection. The APs' operation mode is designed to maximize the minimum SE of the downlink users, while satisfying the sensing requirement to detect the presence of a single target in a certain location.  Relying on the long-term channel state information (CSI), the APs are divided into communication APs (C-APs) and sensing APs (S-AP) to  support downlink communication and sensing operations simultaneously. The main contributions of our paper can be summarized as follows:

\begin{itemize}
\item By leveraging the use-and-then-forget strategy, we derive closed-form expressions for the downlink SE and MRSR to evaluate the performance of the communication and sensing operation, respectively. Then, we formulate the problem of joint AP operation mode selection and power control, considering  per-AP power constraints and a MASR constraint for target detection.

\item  We propose a greedy algorithm for AP operation mode selection. Accordingly, an alternating optimization (AO) algorithm is developed to handle the coupling between the C-AP and R-AP power control coefficients' design.

\item Numerical results show that our proposed greedy AP operation mode selection with optimal power control (GAP-OPC) significantly improves the SE performance of the downlink users for given MASR, compared to the greedy/random operation mode selection with no power control (GAP/RAP-NPC) benchmarks.
 \end{itemize}

\textit{Notation:} We use bold lower case letters to denote vectors. The superscripts $(\cdot)^*$ and $(\cdot)^T$  stand for the conjugate and transpose, respectively;  $\mathbf{I}_N$ denotes the $N\times N$ identity matrix. A zero mean circular symmetric complex Gaussian distribution having variance $\sigma^2$ is denoted by $\mathcal{CN}(0,\sigma^2)$. Finally, $\mathbb{E}\{\cdot\}$ denotes the statistical expectation.

\vspace{-0.7em}
\section{System Model}
\begin{figure}[t]
\centering
\includegraphics[width=0.45\textwidth]{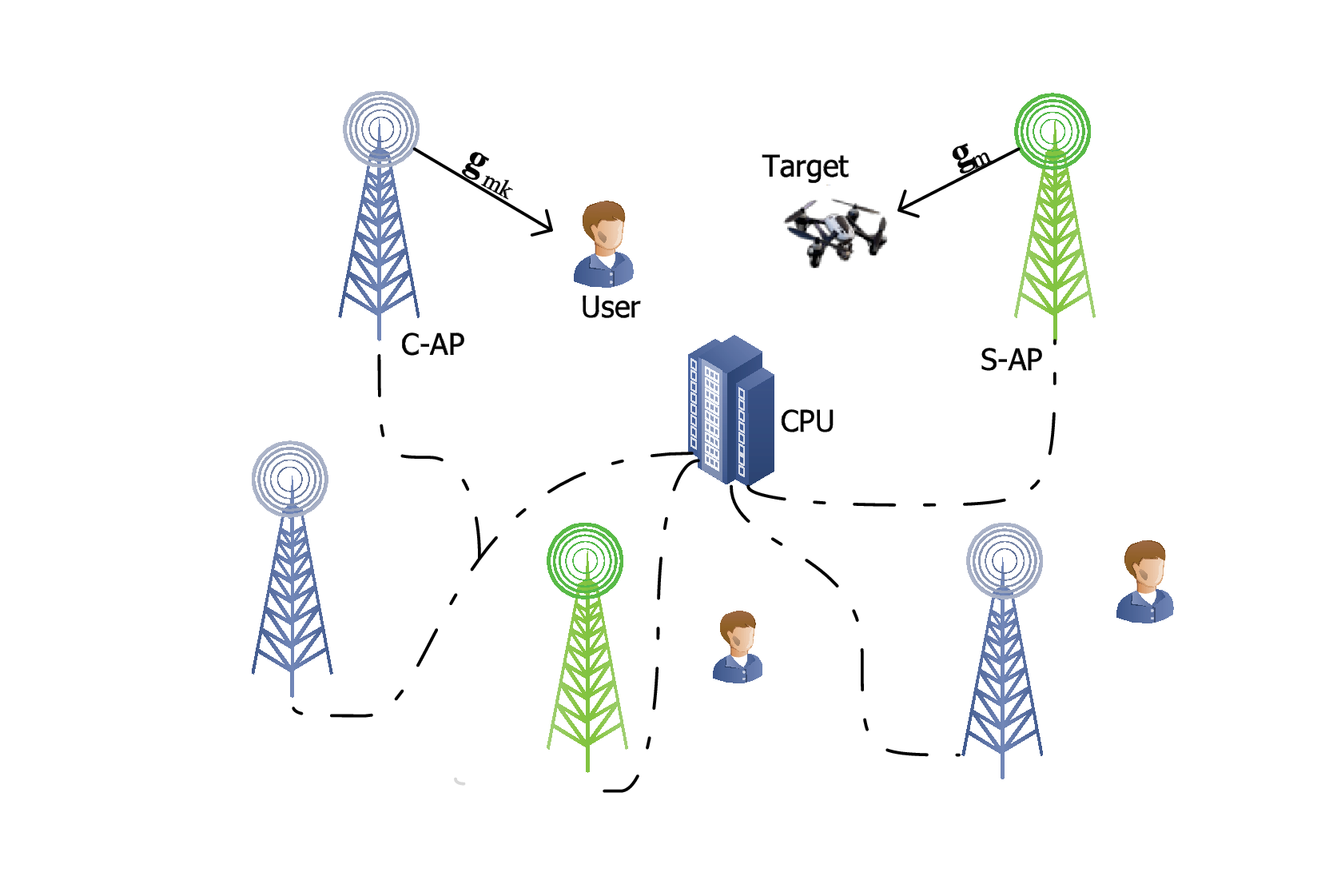}
\vspace{-1em}
\caption{Cell-free massive MIMO ISAC system.}
\vspace{0.3em}
\label{fig:systemmodel}
\end{figure}

We consider a cell-free massive MIMO ISAC system  under time division duplex operation, where $M$ APs serve $K_d$ users in the downlink, while radiating probing signals to a target direction for radar sensing. Each user is equipped with one single antenna, while each AP is equipped with $N$ antennas. All APs and users operate as half-duplex devices. For notational simplicity, we define the sets $\M\triangleq\{1,\ldots,M\}$ and $\K\triangleq \{1,\dots,K_d\}$ as the collections of indices of the APs and users, respectively. As shown in Fig.~\ref{fig:systemmodel},  downlink communication as well as target detection take place simultaneously and over the same frequency band. The AP operation mode selection approach is designed according to the network requirements, determining whether an AP is dedicated to information transmission or radar sensing. The users receive information from a group of the APs, termed as C-APs, while the remaining APs, termed as S-APs, are used for target detection.

\vspace{-0.7em}
\subsection{Channel Model and Uplink Training}
We assume a quasi-static channel model, with each channel coherence interval spanning a duration of $\tau$ symbols. The duration of the training is denoted as $\tau_t$, while the duration of downlink information transfer and target detection is $(\tau-\tau_t)$.

For the sensing channel model, we assume there is a line-of-sight (LOS) path between the target location and each AP, which is a commonly adopted model in the literature~\cite{Behdad:GC:2022,demirhan2023cell}. The LOS channel between AP $m$ and target is given by
\begin{align}~\label{eq:barhkbarH2}
    {\qg}_m &= \qa_N(\phi_{m,t}^a, \phi_{m,t}^e),~\forall m\in \M,
\end{align}
where $\phi_{m,t}^a$, $\phi_{m,t}^e$ denote the azimuth and elevation angles of departure (AoD) from AP $m$ towards the target. Moreover, the $q$-th entry of the array response vector $\qa_N(\phi_{m,t}^a, \phi_{m,t}^e)\in\mathbb{C}^{N\times 1}$, is given by
\vspace{-0.5em}
\begin{align}~\label{eq:LoSarray}
    [\qa_N(\phi_{m,t}^a, \phi_{m,t}^e)]_{q} &\!=\! \frac{1}{\sqrt{N}}\exp\Big( j2\pi\frac{d}{\lambda}  (q\!-\!1) \sin \phi_{m,t}^e\sin \phi_{m,t}^a \Big),
    \end{align}
where $d$ and $\lambda$ denote the AP antenna spacing and carrier wavelength, respectively.

The channel vector between the $m$-th AP and $k$-th user is modeled as $\mathbf{g}_{mk}= \sqrt{\beta_{mk}} \qh_{mk}$,
where $\beta_{mk} $ is the large scale fading coefficients, and $\qh_{mk} \in \mathbb{C}^{N \times 1}$ is the small-scale fading vector, whose elements are independent and identically distributed $\mathcal{CN} (0, 1)$ random variables~\cite{Hien:cellfree}.

An uplink training process is implemented to acquire the local CSI between each AP and all users.
In each coherence block of length $\tau$, all users are assumed to transmit their pairwisely orthogonal pilot sequence of length $\tau_t$ to all APs, which requires $\tau_t\geq K_d$. At AP $m$, $\qg_{mk}$ is estimated by using the received pilot signals and applying the minimum mean-square error (MMSE) estimation technique. By following~\cite{Hien:cellfree}, the MMSE estimate  $\hat{\qg}_{mk}$ of $\qg_{mk}$ is obtained as $\hat{\qg}_{mk}  \sim \mathcal{CN}\left(\boldsymbol{0}, \gamma_{mk} \qI_N \right)$, where
\begin{align}
\gamma_{mk}
&=\frac { \tau_{t} \rho_{t}\beta _{mk}^{2}}
{ \tau_{t} \rho_{t}\beta _{mk}+1},
\end{align}
while $\rho_{t}$ represents the normalized transmit power of each pilot symbol.

\vspace{-0.8em}
\subsection{Data and Probing Signal Transmission}
AP operation mode selection is performed by considering large-scale fading effects and relying on the statistical CSI, obtained during the training phase. The binary variables used to indicate the operation mode for each AP $m$ are defined as
\begin{align}
a_{m} =
 \begin{cases} 1, &\text{if AP $m$ operates as C-AP}\\
  0,  &\text{if AP $m$ operates as S-AP.}
 \end{cases}
\end{align}

The transmission phase comprises information transmission from C-APs to users and probing signal transmission from S-APs to the target. Let $\mathbf{x}_{c,m}$ and $\mathbf{x}_{r,m}$ denote the data and probing signals, respectively, where $\Ex\big\{\vert\qx_{c,m} \vert^2\big\}=1$ and $\Ex\big\{\vert\qx_{r,m} \vert^2\big\}=1$. The signal vector transmitted from AP $m$ can be expressed as
\begin{align}
\mathbf{x}_m={a_{m}} {\qx}_{c,m}+({1-a_{m}}){\qx_{r,m}}.
\end{align}

The power control coefficients at AP $m$ are chosen to satisfy the power constraint at each S-AP and C-AP, respectively, i.e.,
\begin{align}
{a_{m}} \Ex\{\|\mathbf{x}_{c,m}\|^2\}+({1-a_{m}})\Ex\{\|\mathbf{x}_{r,m}\|^2\}\leq \rho,
\end{align}
where $\rho$ denotes the maximum normalized downlink power.

The transmit signal for communication at the $m$-th C-AP can be expressed as $\mathbf{x}_{c,m}= \sum_{k\in\K} \sqrt{{\eta}_{mk}\rho} \tmkcom x_{c,k}$, where  $\eta_{mk}$ represents the downlink power control coefficient at the $m$-th C-AP during communication, while $\tmkcom\in\mathbb{C}^{N\times 1}$ and $x_{c,k}$ denote the precoding vector and intended signal for user $k$,  respectively.

Moreover,  the probing signal transmitted by the $m$-th S-AP can be expressed as $\qx_{r,m}=\sqrt{\eta_m \rho} \tmsen x_{r}$, where $\eta_m$ denotes the power control coefficient at S-AP $m$,  while $\tmsen\in\mathbb{C}^{N\times 1}$ denotes the beamforming vector for sensing and $x_{r}$ is the sensing symbol.

We note that in the absence of communication users, conjugate sensing beamforming solution becomes optimal, as it directly maximizes the sensing SNR~\cite{Behdad:GC:2022,demirhan2023cell}. Therefore, to detect the presence of a target in a certain location, we design the sensing beamforming vector at S-AP $m$ as
\begin{align}~\label{eq:tmsec}
\tmsen = \qa_N(\phi_{m,t}^a, \phi_{m,t}^e).
\end{align}
Furthermore, the conjugate precoder is employed at the C-AP, particularly due to its advantages including low computational complexity, ease of analysis, and reasonable performance, as shown in~\cite{Marzetta:TWC:2010,sutton2021hardening}. Hence, $\tmkcom$ is given by
\begin{align}~\label{eq:tmkcom}
\tmkcom = \hat{\qg}_{mk}^*.
\end{align}

\vspace{-0.8em}
\subsection{Sensing Operation and MASR}
For a given channel realization, the average spatial power
pattern for sensing is defined as
\begin{align}~\label{eq:pavet}
P^{\ave}(\phi_{m,t}^a, \phi_{m,t}^e)&=\Ex\Bigg\{
\sum_{m\in\M}|\qa_N^H(\phi_{m,t}^a, \phi_{m,t}^e)\qx_{m}|^2\Bigg\}
\nonumber\\
&\hspace{-6em}=\rho\sum_{m\in\M}{a_{m}} \Ex\Bigg\{\bigg|\sum_{k\in\K}\sqrt{{\eta}_{mk}}\qa_N^H(\phi_{m,t}^a, \phi_{m,t}^e)   \tmkcom \bigg\vert^2\Bigg\}\nonumber\\
&\hspace{-5em}+\rho\sum_{m\in\M}(1-a_{m})\eta_m \Ex\Big\{\big\vert \qa_N^H(\phi_{m,t}^a, \phi_{m,t}^e)\tmsen \big|^2\Big\},
\end{align}
where the expectation is taken over the transmitted signals, assuming that the information signal and probing signals are independent zero-mean Gaussian distributed.

\begin{proposition}
The average spatial power
pattern for sensing is given by
$P^{\ave}(\phi_{m,t}^a, \phi_{m,t}^e) =P^{\ave}_{\Com}(\phi_{m,t}^a, \phi_{m,t}^e)+P^{\ave}_{\sen}(\phi_{m,t}^a, \phi_{m,t}^e)$, where
\begin{subequations}~\label{eq:pavet:sen:com}
\begin{align}
P^{\ave}_{\Com}(\phi_{m,t}^a, \phi_{m,t}^e)
&=\rho\sum_{m\in\M}\sum_{k\in\K}
a_{m}
{\eta}_{mk}
\gamma_{mk},\\
P^{\ave}_{\sen}(\phi_{m,t}^a, \phi_{m,t}^e)&=\rho\sum_{m\in\M}(1-a_{m})\eta_m.
\end{align}
\end{subequations}
\end{proposition}

\textit{Proof:} By invoking~\eqref{eq:tmsec} and~\eqref{eq:tmkcom}, and then
taking the expectation of~\eqref{eq:pavet} over $\tmkcom$, the desired result is obtained. $\hspace{4em}\blacksquare$

We would like $P^{\ave}{\Com}(\phi{m,t}^a, \phi_{m,t}^e)$, $\forall \phi_{m,t}^a, \phi_{m,t}^e$, to be as small as possible to confine the pattern distortion. For illuminating a target angle $(\phi_{m,t}^a, \phi_{m,t}^e)$, it is desirable that the mainlobe level $P^{\ave}_{\sen}(\phi_{m,t}^a, \phi_{m,t}^e)$ is higher than $P^{\ave}_{\Com}(\phi_{m,t}^a, \phi_{m,t}^e)$ by a certain minimum sensing level $\kappa$, which is referred
to as the MASR:
\begin{align}~\label{eq:MASR}
   \mathrm{MASR}(\aaa,  \ETAmk, \ETAm) &= \frac{P^{\ave}_{\sen}(\phi_{m,t}^a, \phi_{m,t}^e)}{P^{\ave}_{\Com}(\phi_{m,t}^a, \phi_{m,t}^e)}
   \\
   &\hspace{0em}=\frac{\sum_{m\in\M}(1-a_{m})\eta_m}
   {\sum_{m\in\M}\sum_{k\in\K}
a_{m}
{\eta}_{mk}
\gamma_{mk}} \geq \kappa, \nonumber
\end{align}
where $\aaa = \{a_1,\ldots,a_M\}$, $\ETAmk =\{\eta_{m1},\ldots,\eta_{mK_d}\}$, $\forall m\in\M$, and $\ETAm=\{\eta_1,\ldots,\eta_{M}\}$.
\vspace{-0.5em}
\subsection{Communication Operation and Downlink SE}
The received signal at $k$-th user can be expressed as
\begin{align}\label{eq:yk}
{y}_k&={\sum_{m\in\mathcal{M}} a_m \sqrt{\rho \eta_{m k}} \mathbf{g}^{T}_{m k} \tmkcom x_{c,k}}
\nonumber\\
&\hspace{2em}+{ \sum_{m\in\mathcal{M}} \sum_{k'\in\K \setminus k} a_m \sqrt{\rho \eta_{m k'}} \mathbf{g}^{T}_{mk} \tmkpcom  x_{c,k'}}\nonumber\\
&\hspace{2em}+{\sum_{m\in\mathcal{M}}\left(1-a_m\right) \sqrt{\rho \eta_m}\qg^{T}_{mk}\tmsen  x_{r}}+{n}_k,
\end{align}
where the second term is the inter-user interference, the third term represents the interference from S-APs, and $n_k\sim\mathcal{CN}(0,\sigma^2_n)$ denotes the additive white Gaussian noise at the user $k$.

\begin{proposition}~\label{Prop:SE:UE}
    With conjugate precoding at the APs for downlink communication, the achievable downlink SE of user $k$, can be expressed as $\SEk=\Big(1-\frac{\tau_{\mathrm {p}}}{\tau}\Big)\log _{2}\big(1+\SINR_k(\aaa,  \ETAmk, \ETAm)\big)$, where $\SINR_k(\aaa,  \ETAmk, \ETAm)$ is given by~\eqref{eq:SINE} at the top of the next page.
\vspace{0.8em}
\begin{figure*}
\vspace{0.7em}
    \begin{align}~\label{eq:SINE}
    &\SINR_k(\aaa,  \ETAmk, \ETAm)=
    \!\frac{
                 \rho N^2\big(\sum_{m\in\mathcal{M}} \! a_m  \eta_{m k}^{1/2} \gamma_{mk}\big)^2
                 }
                 {
                 \rho N \!\!\sum_{m\in\mathcal{M}} \! \sum_{k'\in\K}\!  a_m \eta_{m k'} \gamma_{mk'} \beta_{mk}
                \! +\!\rho \!\sum_{m\in\M}(1\!-\!a_m){\eta_m }  \beta_{mk}  \!+ \! 1}\!.
\end{align}
\hrulefill
\vspace{-2mm}
\end{figure*}
\end{proposition}

\textit{Proof:} See Appendix~\ref{Proof:Prop:SE:UE}. $\hspace{14em}\blacksquare$

\vspace{-0.1em}
\section{Proposed Design Problem and Solution}
In this section, we formulate and solve the AP mode selection to maximize the minimum SE. More specifically, we aim to optimize the AP operation mode selection vector ($\qa$) and power control coefficients ($\ETAmk, \ETAm$) to maximize the minimum per-user SE subject to a prescribed MASR level for the target detection and transmit power constraints at the APs. The optimization problem is then formulated as
\begin{subequations}\label{P:MASR}
	\begin{align}
		\text{\textbf{(P1):}}~\underset{\aaa, \ETAmk, \ETAm}{\max} & \min_{k\in\K}\,\, \hspace{0.5em}
		\SINR_k (\aaa,  \ETAmk, \ETAm)
		\\
		\mathrm{s.t.} \,\,
		\hspace{0.5em}&  \mathrm{MASR}(\aaa,  \ETAmk, \ETAm) \geq \kappa,\label{eq:MASR:ct1}\\
			&a_m\sum_{k\in\K}{\eta_{mk}}\gamma_{mk}\leq \frac{1}{N},
        ~\forall m\in\M,\label{eq:etamk:ct1}\\
       &{\eta_m}\leq 1-a_m,~\forall m\in\M,\label{eq:etam:ct1}\\
		& a_m \in\{ 0,1\}.\label{eq:am:binary}
		\end{align}
\end{subequations}

Problem (\textbf{P1}) is  a challenging combinatorial problem. Therefore, for AP operation mode selection, we only focus on  a heuristic greedy method which simplifies the computation, while providing a significantly successful monitoring performance gain.

\vspace{-0.2em}
\subsection{AP Operation Mode Selection with Fixed Power Control}
Let $\mathcal{A}_{\sen}$ and $\mathcal{A}_{\Com}$ denote the sets containing the indices of APs operating as radar, i.e., APs with $a_m=0$, and APs operating in communication mode, i.e., APs with $a_m=1$, respectively.  In addition, $\mathrm{MASR}(\mathcal{A}_{\sen}, \mathcal{A}_{\Com})$ and $\SINR_k(\mathcal{A}_{\sen}, \mathcal{A}_{\Com})$ underline the dependence of the sensing MASR  and received SINR of the $k$-th user on the different choices of AP mode selections. Our greedy algorithm of AP mode selection is shown in \textbf{Algorithm~\ref{alg:Grreedy}}. To guarantee the sensing MASR requirement, all APs are initially assigned for sensing operation, i.e., $\mathcal{A}_{\sen}=\mathcal{M}$ and $\mathcal{A}_{\Com}=\emptyset$. Then, in each iteration, one AP switches into communication operation mode for maximizing the minimum of SE (or equivalently SINR in~\eqref{eq:SINE}),  while the minimum MARS required for target sensing is guaranteed. This process continues until there is no more improvement in the minimum SINR among all users.

\begin{algorithm}[!t]
\caption{Greedy AP Operation Mode Selection}
\begin{algorithmic}[1]
\label{alg:Grreedy}
\STATE
\textbf{Initialize}: Set  $\mathcal{A}_{\Com}=\emptyset$ and $\mathcal{A}_{\sen}=\mathcal{M}$. Set iteration index $i=0$.
\STATE Calculate $\Pi^{\star}[i]=  \min_{k\in\K}\SE_k({A}_{\sen}, \mathcal{A}_{\Com})$
\REPEAT
\FORALL{$m \in \mathcal{A}_{\sen}$}
\STATE Set $\mathcal{A}_{s}=\mathcal{A}_{\sen} \setminus m$.\\
\IF{ $\mathrm{MASR}(\mathcal{A}_{s},\mathcal{A}_{\Com} \bigcup m)\!\geq\!\kappa$ }
\STATE  Calculate $\Pi_m= \min_{k\in\K}\SINR_k( \mathcal{A}_{s},\mathcal{A}_{\Com} \bigcup m)$\\
 \ELSE
\STATE Set  $\Pi_m=0$
\ENDIF
\ENDFOR
\STATE Set $\Pi^\star[i+1]= \underset{m\in\mathcal{A}_{\Com}} \max \,\,\Pi_m$\\
\STATE $e=|\Pi^\star[i+1]- \Pi^\star[i]|$
\IF{$e \geq e_{\min}$ }
\STATE {Update $\mathcal{A}_{\Com}\!=\!\{\mathcal{A}_{\Com}\bigcup m^{\star}\}$ and $\mathcal{A}_{\sen}\!=\!\mathcal{A}_{\sen}\!\setminus\! m^{\star}$}
\ENDIF
\STATE {Set $i=i+1$}
\UNTIL{ $e < e_{\min}$ }
\RETURN $\mathcal{A}_{\sen}$ and $\mathcal{A}_{\Com}$, i.e., the indices of APs operating in radar mode and communication mode, respectively.
\end{algorithmic}
\end{algorithm}
\setlength{\textfloatsep}{0.2cm}
\vspace{-0.2em}
\subsection {Power Control}
For a given AP mode selection, the optimization problem~\eqref{P:MASR} reduces to the power control problem, given by
\begin{subequations}\label{P:MASR:power:p2}
\begin{align}
		\text{\textbf{(P2):}}~\underset{ \ETAmk, \ETAm}{\max}\,\, \hspace{0.5em}&
	\min_{k\in\K}\,\, \hspace{0.5em}
		\SINR_k ( \ETAmk, \ETAm)
             \\
		&\hspace{-2em}\mathrm{s.t.} \,\,
		\hspace{0.5em} \eqref{eq:MASR:ct1}-\eqref{eq:am:binary}.
\end{align}
\end{subequations}

Problem (\textbf{P2}) is a non-convex optimization problem due the to non-convex objective function and constraints. Since the variables $\ETAmk$ and $\ETAm$ are coupled in both the objective and MASR constraint, it is difficult to simultaneously optimize them. Therefore, we propose an AO algorithm to jointly optimize $\ETAmk$ and $\ETAm$ in two sub-problems.

Firstly, for a given $\ETAm$, we formulate the sub-problem for optimizing $\ETAmk$ as
\begin{subequations}\label{P:MASR:power:p2-1}
\begin{align}
		\text{\textbf{(P2-1):}}~\underset{ \ETAmk}{\max}\,\, \hspace{0.5em}&
	\min_{k\in\K}\,\, \hspace{0.5em}
		\SINR_k ( \ETAmk)
             \\
		&\hspace{-2em}\mathrm{s.t.} \,\,
		\hspace{0.5em} \eqref{eq:MASR:ct1},\eqref{eq:etamk:ct1}.
\end{align}
\end{subequations}

By introducing the slack variables $\theta_{mk} = \eta_{mk}^{\frac{1}{2}}$ and $ \upsilon_m$, we reformulate (\textbf{P2-1}) as
\begin{subequations}\label{P:MASR:power:p2-1-1}
\begin{align}
	\text{\textbf{(P2-2):}}~\underset{ \ETAmk, t}{\max}\,\, \hspace{0.5em}
	t\\
 &\hspace{-2em}\mathrm{s.t.} \,\,\hspace{0.5em} \!\frac{
                 \big(\sum_{m\in\mathcal{M}} \! a_m  \theta_{m k} \gamma_{mk}\big)^2
                 }
                 {
                 \frac{1}{N} \!\!\sum_{m\in\mathcal{M}} a_m \beta_{mk}  \upsilon_m^2
                \! +\varphi_k}\geq t,~\forall k\in\K\!
             \\
             &\frac{\sum_{m\in\M}(1-a_{m})\eta_m}
   {\sum_{m\in\M}
a_{m}
\upsilon_m^2} \geq \kappa,\label{eq:ct2:P2-2}\\
        &\sum_{k'\in\K}\!  a_m \theta_{m k'}^2 \leq \upsilon_m^2,~\forall m\in\M\label{eq:ct3:P2-2}\\
        &0 \leq a_m \upsilon_m^2\leq \frac{1}{N},
        ~\forall m\in\M,\label{eq:etamk:ct1:new}\\
        &\theta_{mk}\geq 0, \forall m\in\M,\forall k\in\K\label{eq:ct5:P2-2},
\end{align}
\end{subequations}
where $\varphi_k \overset{\Delta}{=} \frac{1}{N^2} \!\sum_{m\in\M}(1\!-\!a_m){\eta_m }  \beta_{mk}  \!+ \frac{1}{\rho N^2}$ and and $\upsilon_m^2 \overset{\Delta}{=} \sum_{k\in\K}\!  a_m \eta_{m k}$. The equivalence between~\eqref{P:MASR:power:p2-1-1} and~\eqref{P:MASR:power:p2-1} follows directly from the fact that the second constraint in~\eqref{P:MASR:power:p2-1-1} holds with equality at the optimum. Problem ($\textbf{P2-2}$) can be reformulated as a second-order cone program (SOCP). More precisely, for given $t$, we have
\vspace{0.2em}
\begin{align}\label{P:MASR:power:p2-1-2}
	\text{\textbf{(P2-3):}}~\underset{ \ETAmk}{\max}\,\, \hspace{0.5em}
	t\\
 &\hspace{-2em}\mathrm{s.t.} \,\,\hspace{0.5em}\Vert \qv_k\Vert \leq \frac{1}{\sqrt{t}}
                 \sum_{m\in\mathcal{M}} \! a_m  \theta_{m k} \gamma_{mk},~\forall k\in\K,
             \\
        &~\eqref{eq:ct2:P2-2}-~\eqref{eq:ct5:P2-2},
\end{align}
where $\qv_k = [\qv_{k1}^T, \sqrt{\varphi_k}]^T$, with $\qv_{k1} = \big[\sqrt{\frac{\beta_{1k}}{N}}\upsilon_1,\ldots,\sqrt{\frac{\beta_{Mk}}{N}}\upsilon_M\big]^T$. The first constraint represents a second order cone and thus ($\textbf{P2-3}$) is a standard SOCP, which is a convex problem. The bisection search method is exploited to find the optimal solution, in each step solving a sequence of convex feasibility problem. This bisection based search method is summarized in \textbf{Algorithm 2}.

\begin{algorithm}[t]\label{Alg:PA}
\caption{Bisection Method for Power Control }
\begin{algorithmic}[1]
\STATE Initialization of $t_{\min}$ and $t_{\max}$, where $t_{\min}$ and $t_{\max}$ define a range of relevant values of the objective function in~\eqref{P:MASR:power:p2-1}. Initial line-search accuracy  $\epsilon$.
\REPEAT
    \STATE Set $t:=\frac{t_{\min}+t_{\max}}{2}$. Solve the following convex feasibility program
    \begin{align}\label{eq:Feasibility}
    \begin{cases} \hspace {0.0 cm}
	&\Vert \qv_k\Vert \leq \frac{1}{\sqrt{t}}
                 \big(\sum_{m\in\mathcal{M}} \! a_m  \theta_{m k} \gamma_{mk}\big),~\forall k\in\K\!.
             \\
 &\hspace{0.5em}
   {\sum_{m\in\M}
a_{m}
\upsilon_m^2} \leq \frac{1}{\kappa}{\sum_{m\in\M}(1-a_{m})\eta_m},\\
        &\sum_{k'\in\K}\!  a_m \theta_{m k'}^2 \leq \upsilon_m^2,~\forall m\in\M\\
        &0 \leq a_m \upsilon_m\leq \frac{1}{\sqrt{N}},
        ~\forall m\in\M,\\
        &\theta_{mk}\geq 0, \forall m\in\M,\forall k\in\K,
 \end{cases}
\end{align}
\STATE If problem~\eqref{eq:Feasibility} is feasible, then set $t_{\min}:=t$, else set $t_{\max} :=t$.
\UNTIL{ $t_{\max}-t_{\min}<\epsilon$.}
\end{algorithmic}
\end{algorithm}
\setlength{\textfloatsep}{0.1cm}

Secondly, when $\ETAmk$ is fixed, the sub-problem for optimizing $\ETAm$ can be expressed as
\vspace{0.2em}
\begin{subequations}\label{P:MASR:power:p2-4}
\begin{align}
		\text{\textbf{(P2-4):}}~\underset{ \ETAm}{\max}\,\, \hspace{0.5em}&
	\min_{k\in\K}\,\, \hspace{0.5em}
		\SINR_k ( \ETAm)
             \\
		&\hspace{-2em}\mathrm{s.t.} \,\,
		\hspace{0.5em} \eqref{eq:MASR:ct1},\eqref{eq:etam:ct1}.
\end{align}
\end{subequations}

By introducing a new slack variable $\varrho$, we can reformulate the optimization problem as
\vspace{0.2em}
\begin{subequations}\label{P:MASR:power:p2-4}
\begin{align}
	\text{\textbf{(P2-5):}}~\underset{ \ETAm, \varrho}{\max}\,\, \hspace{0.5em}
	\varrho\\
 &\hspace{-2em}\mathrm{s.t.} \,\,\hspace{0.5em} \!
                 \frac{\!\sum_{m\in\M}(1\!-\!a_m)
                 {\eta_m }  \beta_{mk}  \!+ \phi_k}{\big(N\sum_{m\in\mathcal{M}} \! a_m  \theta_{m k} \gamma_{mk}\big)^2} \leq                 \frac{1}{\varrho},~\forall k\in\K\!
             \\
             &\sum_{m\in\M}(1-a_{m})\eta_m
   \geq \kappa \sum_{m\in\M}
a_{m} \upsilon_m^2,\label{eq:ct2:P2-2}\\
&{\eta_m}\leq 1-a_m,~\forall m\in\M,\label{eq:etam:ct1}
\end{align}
\end{subequations}
where $\phi_k\overset{\Delta}{=}\frac{1}{N} \!\!\sum_{m\in\mathcal{M}} a_m \beta_{mk}  \upsilon_m^2 \! + \frac{1}{\rho N^2}$. Now, for a fixed $\varrho$, all inequalities involved in (\textbf{P2-5}) are linear,  hence the solution to the optimization problem can be obtained by harnessing a line-search over $\varrho$ to find the maximal feasible value. Therefore, we can apply the bisection method in \textbf{Algorithm 2} to solve~\eqref{P:MASR:power:p2-4}, where $t_{\min}$, $t_{\max}$, and the feasibility problem~\eqref{eq:Feasibility} are replaced with $\varrho_{\min}$, $\varrho_{\max}$ and problem~\eqref{P:MASR:power:p2-4}, respectively. We summarize the overall AO algorithm in \textbf{Algorithm~3}.

\vspace{-1.9em}
~\subsection{Complexity Analysis}
Here, we provide the computational complexity of Algorithm 3, which involves a SOCP problem in~\eqref{P:MASR:power:p2-1-2} and a linear-search problem~\eqref{P:MASR:power:p2-4} at each iteration. In order to solve a SOCP, the iterative bisection search method requires $\mathcal{O}(n_{v_1}^2 n_c)$ arithmetic operations, where $n_{v_1}$ is the number of optimization variables and $n_c$ is the total number of SOC constraints~\cite{Boyd04}. Moreover, the total number of iterations required is $\log_2\big(\frac{t_{\max}-t_{\min}}{\epsilon}\big)$. In~\eqref{P:MASR:power:p2-1-2}, the total number of variables is $n_{v_1}=MK_d$ and there are $n_v=K_d$ SOC constraints. Therefore, the per-iteration computational complexity for solving ~\eqref{P:MASR:power:p2-1-2} is $\log_2\big(\frac{t_{\max}-t_{\min}}{\epsilon}\big)\mathcal{O}(MK_d^3)$. Problem~\eqref{P:MASR:power:p2-4} involves $n_{v_2}=M$ scalar-value variables and $n_{c_2}=M+K_d+1$ linear constraints. According to~\cite{Vincent:TWC:2017}, the per-iteration cost to solve~\eqref{P:MASR:power:p2-4} is $\mathcal{O}\big( (n_{c_2}+n_{v_2}) n_{v_2}^2 n_{c_2}^{0.5} \big)$.
\begin{algorithm}[t]\label{Alg:AO}
\caption{AO Algorithm for Problem \textbf{P2}}
\begin{algorithmic}[1]
\STATE Initialize a feasible initial point, $(\ETAm)^{(0)}$ and $(\ETAm)^{(0)}$.
\STATE Set the iteration number $n=1$.
\REPEAT
\STATE Determine $(\ETAmk)^{(n)}$ by using \textbf{Algorithm 2}.
\STATE Compute $(\ETAm)^{(n)}$ by solving~\eqref{P:MASR:power:p2-4}.
\STATE Set $n=n+1$.
\UNTIL{ some stopping criterion is satisfied. }
\end{algorithmic}
\end{algorithm}
\setlength{\textfloatsep}{0.05cm}

\vspace{-0.1em}
\section{Numerical Results}
We assume that the $M$ APs and $K_d$ users are uniformly distributed at random within a square of size $D \times D~\text{km}^2$, whose edges are wrapped around to avoid the boundary effects. The large-scale fading coefficient $\beta_{mk}$ models the path loss and shadow fading, according to $\beta_{mk} = \mathrm{PL}_{mk} 10 ^{\frac{\sigma_{sh} z_{mk}}{10}}$, where $\mathrm{PL}_{mk}$ represents the path loss, and $10 ^{\frac{\sigma_{sh} z_{mk}}{10}}$ represents the shadow fading with the standard deviation $\sigma_{sh}$, and $z_{mk} \sim \mathcal{CN}(0,1)$. We use the three-slope model for the path-loss $\mathrm{PL}_{mk}$ (in dB) as
\begin{align*}
\mathrm{PL}_{mk} \!=\!\! \left\{ \begin{array}{ll}
\!\!\!\!\!-L\!-35\log_{10}(d_{mk})         & d_{mk}>d_1,  \\
\!\!\!\!\!-L\!-15\log_{10}(d_{1})\!-20\log_{10}(d_{mk})            &  d_0<d_{mk}\leq d_1\\
\!\!\!\!\!-L\!-15\log_{10}(d_{1})\!-20\log_{10}(d_{0}) &  d_{mk}\leq d_0.\end{array} \right.
\end{align*}
where $L$ is a constant depending on the carrier frequency, the
user and AP heights, given in~\cite{Hien:cellfree}. We  further use the correlated shadowing model for $d_{mk}>d_1$~\cite{Hien:cellfree}. Here, we choose $\sigma_{sh}=8$ dB, $D=0.5$ km, $d_1=50$ m, and $d_0=10$ m. We further set the noise power $\Sn=-108$ dBm. Let $\tilde{\rho} =1$ W and $\tilde{\rho}_t =0.25$ W be the maximum transmit power of the APs and uplink training pilots, respectively. The normalized maximum transmit powers ${\rho}$ and ${\rho}_t$ are calculated by dividing these powers by the noise power $\Sn$.

In the absence of power control, termed as no power control (NPC) design, both C-APs and S-APs transmit at full power. With NPC, the power coefficients at the $m$-th C-AP are the same and $\etamk = \big(N\sum_{k\in\K} \gammk \big)^{-1} $, $\forall k\in\K$. Moreover $\etam = 1$, $\forall m\in\M$. For comparison, two benchmark system designs are studied: 1) Random AP mode selection with NPC (RAP-NPC), and 2) Greedy AP mode selection with NPC (GAP-NPC).
\begin{figure}[t]
	\vspace{-0.5em}
	\centering
	\includegraphics[width=0.42\textwidth]{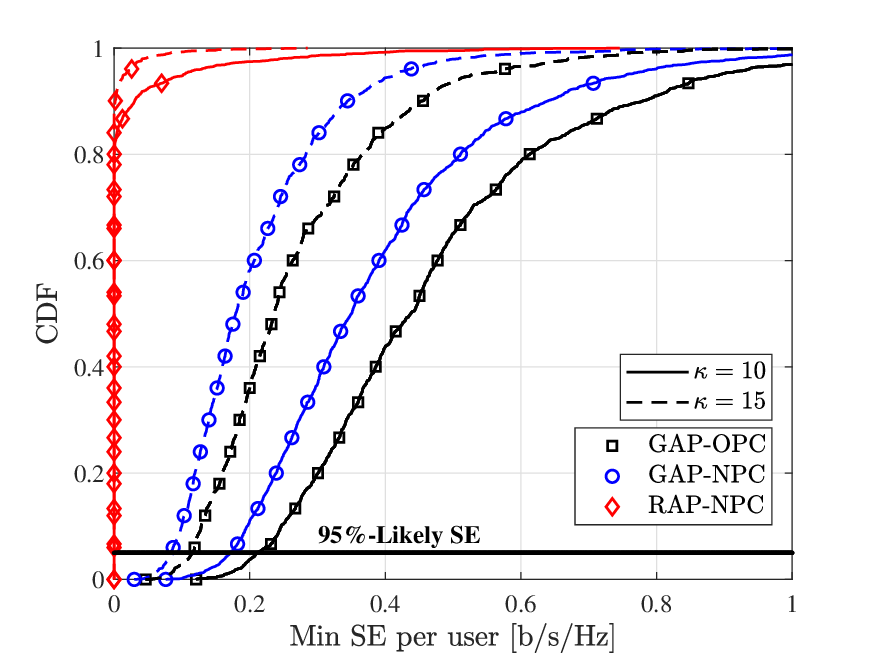}
 \vspace{-0.6em}
	\caption{The CDF of the minimum per user SE for different values of $\kappa$ ($M=80$, $N=3$, $K_d=5$).}
	\vspace{-1.2em}
	\label{fig:Fig2}
\end{figure}

Figure~\ref{fig:Fig2} shows the cumulative distribution function (CDF) of the minimum per-user SE for two different values of $\kappa$. For a fair comparison, the achievable SE of the RAP-NPC is set to zero, when the MASR target value is not satisfied. It is noteworthy that the $95\%$-likely minimum per-user SE increases, when $\kappa$ decreases from $15$ to $10$. This behavior can be explained by the fact that a great number of APs are assigned for downlink communication by reducing $\kappa$. Furthermore, by applying power control, the $95\%$-likely SE is improved by $50\%$ (for $\kappa=15$) and $35\%$ (for $\kappa=10$) compared to the GAP-NPC design.

\begin{figure}[t]
	\vspace{-1em}
	\centering
	\includegraphics[width=0.42\textwidth]{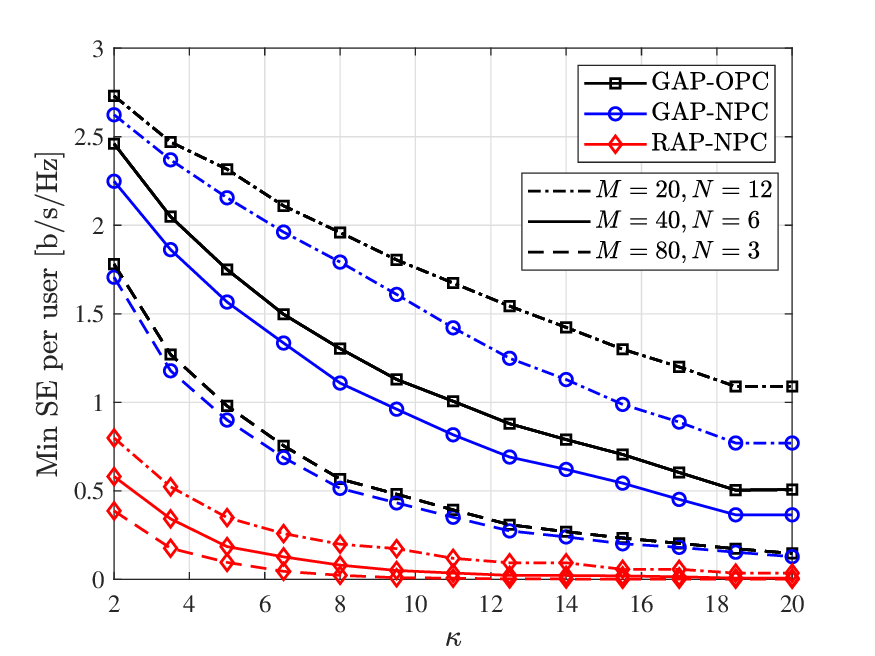}
 \vspace{-0.6em}
	\caption{Average minimum SE versus MASR level ($K_d=5$, $NM=240$).}
	\vspace{0em}
	\label{fig:Fig3}
\end{figure}

Figure~\ref{fig:Fig3} shows the minimum SE of the downlink communication system versus the MASR requirement of the radar system, $\kappa$. In this figure, we assume that a fixed number of service antennas, i.e., $MN=240$, is utilized to support both communication and sensing applications. We observe that by increasing $\kappa$, the achievable downlink SE is decreased since the number of C-APs is decreased. Moreover, while RAP-NPC nearly fails to satisfy the MASR requirements, the GAP scheme with NPC and OPC not only meets the MASR requirements of the sensing operation, but also provides significant SE for the communication system, especially when the number of antennas per each AP is increased from $N=3$ to $N=12$.

\vspace{-0.7em}
\section{Conclusion}
In this paper, we studied the problem of AP operation mode selection and power control design in a cell-free massive MIMO ISAC system, aiming to support multiuser downlink communication and single-target detection. After deriving closed-form expressions for the per-user effective SINR and sensing MASR, we formulated a max-min SE fairness problem. To deal with the complicated non-convex problem, a greedy algorithm, grounded on long-term statistics, was proposed for the AP operation mode design and an AO algorithm was developed for power control design at all APs. Our numerical results highlighted that our proposed GAP-OPC design significantly enhances the downlink SE compared to the GAP-NPC and RAP-NPC benchmarks, while maintaining reliable sensing performance.

\appendices
\vspace{-0.8em}
\section{Proof of Proposition~\ref{Prop:SE:UE}}~\label{Proof:Prop:SE:UE}
In order to apply the use-then-forget technique to derive the downlink SE at user $k$, we rewrite~\eqref{eq:yk} as
\begin{align}\label{eq:yk:utf}
    y_k &=  \mathrm{DS}_k  x_{c,k} \!+\!
    \mathrm{BU}_k x_{c,k}
     \!+\!\sum_{k'\in\K \setminus k}\!\!\!
     \mathrm{IUI}_{kk'}
     x_{c,k'}
    \!+\! \IR_k x_{r} \!+\! n_k,
\end{align}
where
\vspace{-0.1 em}
\begin{subequations}
  \begin{align}
 \mathrm{DS}_k  &\overset{\Delta}{=} \sum_{m\in\mathcal{M}} a_m \sqrt{\rho \eta_{m k}} \Ex\Big\{\mathbf{g}^{T}_{m k} \tmkcom \Big\}
 \\
 \mathrm{BU}_k  &\overset{\Delta}{=}
\sum_{m\in\mathcal{M}} a_m\sqrt{\rho \eta_{mk}}
 \Big( \qg_{mk}^T\qt_k - \Ex\big\{\qg_{mk}^T\qt_k \big\}\Big),
 \\
 \mathrm{IUI}_{kk'} &\overset{\Delta}{=}
 \sum_{m\in\mathcal{M}} \sum_{k'\in\K \setminus k} a_m \sqrt{\rho \eta_{m k'}} \mathbf{g}^{T}_{mk} \tmkpcom,
   \\
  \IR_k &\overset{\Delta}{=} \sum_{m\in\M}(1-a_m) \sqrt{\eta_m \rho}\qg_{mk}^T \tmsen,~\label{eq:IRk}
 \end{align}
\end{subequations}
respectively  represent the strength of the desired signal ($\mathrm{DS}_k$),
the beamforming gain uncertainty ($\mathrm{BU}_k$), interference caused by the $k'$-th user ($\mathrm{IUI}_{kk'}$) and the interference caused by S-APs ($\IR_k$), respectively. By invoking~\eqref{eq:yk:utf}, the achievable downlink SE at the $k$-th user can be expressed as $\SEk=\Big(1-\frac{\tau_{\mathrm {p}}}{\tau}\Big)\log _{2}\big(1 + \SINR_k\big)$, where
\begin{align}~\label{eq:dLSE}
\SINR_k\!=\!\frac{
                 \big\vert  \mathrm{DS}_k  \big\vert^2
                 }
                 {
                 \Ex\Big\{ \big\vert  \mathrm{BU}_k  \big\vert^2\Big\} \!+\!
                 \sum_{k'\in\K \setminus k}
                  \Ex\Big\{ \big\vert \mathrm{IUI}_{kk'} \big\vert^2\Big\}
                  \! +\!  \Ex\Big\{ \big\vert \IR_k \big\vert^2\Big\}  \!+\!  1}\!.
\end{align}

Now, we proceed to the desired signal term as
\begin{align}~\label{eq:DS:ap}
 \mathrm{DS}_k
 &
 =\sqrt{\rho}\Ex\Big\{\sum_{m\in\mathcal{M}} a_m  \eta_{m k}^{1/2}
 (\hat{\qg}_{m k} +\tilde{\qg}_{m k})^{T}\hat{\qg}_{m k}^*\Big\}\nonumber\\
 &
 =\sqrt{\rho}\sum_{m\in\mathcal{M}} a_m  N\eta_{m k}^{1/2} \gamma_{mk},
\end{align}
where we have used the fact that $\hat{\qg}_{m k}$ and  $ \tilde{\qg}_{mk}$ are zero mean and independent.

Noticing that  the variance of a sum of
independent RVs is equal to the sum of the variances, we can derive $\Ex\Big\{ \big\vert  \mathrm{BU}_k  \big\vert^2\Big\}$ as
\begin{align*}
\Ex\Big\{ \big\vert  \mathrm{BU}_k  \big\vert^2\Big\}
 &= \rho\sum_{m\in\mathcal{M}} a_m  \eta_{m k} \Ex\Big\{ \Big\vert\qg_{m k}^{T}\hat{\qg}_{m k}^*-\Ex\Big\{ \qg_{m k}^{T}\hat{\qg}_{m k}^*\Big\}\Big\vert^2\Big\}\nonumber\\
  &\hspace{-2em}= \rho\!\!\sum_{m\in\mathcal{M}}\!\! a_m  \eta_{m k}
 \Big(\Ex\Big\{ \big\vert\qg_{m k}^{T}\hat{\qg}_{m k}^*\big\vert^2\Big\}\!-\!\Big\vert\Ex\Big\{ \qg_{m k}^{T}\hat{\qg}_{m k}^*\Big\}\Big\vert^2\Big\}\Big)\nonumber\\
 &\hspace{-2em}
 = \rho\!\!\sum_{m\in\mathcal{M}} \!\!a_m  \eta_{m k}  \Big(\Ex\Big\{ \big\vert\tilde{\qg}_{m k}^{T}\hat{\qg}_{m k}^*\vert^2\Big\}\!+\!\Ex\Big\{ \vert\vert\hat{\qg}_{m k}\vert\vert^4 \Big\}\!-\! N^2\gamma_{mk}^2\Big).
\end{align*}

By using the fact that $\Ex\Big\{ \big\vert\tilde{\qg}_{m k}^{T}\hat{\qg}_{m k}^*\big\vert^2\Big\}=\Ex\Big\{ \tilde{\qg}_{m k}^{T}\Ex\big\{\hat{\qg}_{m k}^*\hat{\qg}_{m k}^T\big\} \tilde{\qg}_{m k}^*\Big\}=N\gamma_{mk}(\beta_{mk}-\gamma_{mk})$ and  $\Ex\Big\{ \vert\vert\hat{\qg}_{m k}\vert\vert^4 \Big\}=N(N+1)(\gamma_{mk})^2$, we get
\begin{align}~\label{eq:BU:ap}
\Ex\Big\{ \big\vert  \mathrm{BU}_k  \big\vert^2\Big\}
 =\rho N \!\sum_{m\in\mathcal{M}}\! a_m \eta_{m k} \gamma_{mk} \beta_{mk}.
\end{align}

By following the same steps, we can obtain
\begin{align}~\label{eq:IUI:ap}
\Ex\Big\{ \big\vert \mathrm{IUI}_{kk'} \big\vert^2\Big\}=\rho N \sum_{m\in\mathcal{M}} a_m \eta_{m k'} \gamma_{mk'} \beta_{mk}.
\end{align}

Moreover, by substituting~\eqref{eq:tmsec} into~\eqref{eq:IRk} and noticing that $\tmsen (\tmsen)^H = \frac{1}{N} \qI_N$, we get
\begin{align}~\label{eq:IR:ap}
  \Ex\Big\{ \big\vert \IR_k \big\vert^2\Big\}
  &= \sum_{m\in\M}(1-a_m) \Ex\Big\{\vert\qg_{mk}^T \tmsen\vert^2\Big\}
  \nonumber\\
  &=\rho\sum_{m\in\M}{\eta_m }(1-a_m) \beta_{mk}.
\end{align}
To this end, by substituting~\eqref{eq:DS:ap},~\eqref{eq:BU:ap},~\eqref{eq:IUI:ap}, and~\eqref{eq:IR:ap} into~\eqref{eq:dLSE}, the desired result in~\eqref{eq:SINE} is obtained.

\vspace{0em}
\bibliographystyle{IEEEtran}


\end{document}